\documentclass[sigconf]{acmart}
\AtBeginDocument{%
  }
\usepackage{multicol,multirow}
\usepackage{xcolor}

\usepackage{placeins}
\setcopyright{acmlicensed}
\copyrightyear{2018}
\acmYear{2018}
\acmDOI{XXXXXXX.XXXXXXX}
\acmConference[Conference acronym 'XX]{Make sure to enter the correct
  conference title from your rights confirmation email}{June 03--05,
  2018}{Woodstock, NY}
\acmISBN{978-1-4503-XXXX-X/2018/06}

\begin{document}

\title{What Makes a Good Semantic ID for Generative Recommendation? A Reproducibility Study}

\author{Yufei Chen}
\authornote{Equal contribution.}
\affiliation{
  \institution{Shandong University}
  \city{Jinan}
  \country{China}
}
\email{cyf200409@gmail.com}

\author{Junchen Fu}
\authornotemark[1]
\affiliation{
  \institution{University of Glasgow}
  \city{Glasgow}
  \country{United Kingdom}
}
\email{j.fu.3@research.gla.ac.uk}

\author{Jujia Zhao}
\affiliation{
  \institution{Leiden University}
  \city{Leiden}
  \country{Netherlands}
}
\email{zhao.jujia.0913@gmail.com}

\author{Yukun Zhao}
\affiliation{
  \institution{Shandong University}
  \city{Jinan}
  \country{China}
}
\email{zhaoyukun@sdu.edu.cn}

\author{Zhaochun Ren}
\affiliation{
  \institution{Leiden University}
  \city{Leiden}
  \country{Netherlands}
}
\email{z.ren@liacs.leidenuniv.nl}

\makeatletter
\g@addto@macro\addresses{\@@authornotemark{2}} 
\g@addto@macro\@authornotes{\footnotetext[2]{Corresponding author.}}

\begin{abstract}
Generative recommendation has emerged as an active research direction, where items are commonly represented by semantic IDs (SIDs): discrete codes generated token by token. Despite strong empirical results, SID designs vary widely in construction strategy, codebook organization, and code length, making their true impact on recommendation performance unclear.

We conduct a large-scale reproducibility study to systematically investigate the impact of semantic ID design on generative recommendation under a unified experimental framework. We focus on a fundamental question: \emph{What makes a good semantic ID for generative recommendation?} To answer this question, we examine four aspects: the relative effectiveness of different semantic ID designs, the connection between codebook utilization and recommendation quality, the effect of semantic code length, and the influence of semantic ID design on local item semantic preservation. Through a unified evaluation and additional cross-dataset controlled analyses, we find that the effects of SID design are largely non-monotonic: no single SID design is universally best, and commonly used RQ-VAE- and OPQ-based designs can behave inconsistently across datasets. The method with the most balanced first-level codebook is not consistently the best recommender, showing that utilization is diagnostic but insufficient. Scaling either the generative backbone or the SID length is also not always beneficial. Finally, semantic-neighborhood analysis reveals that no single SID design dominates all notions of local semantic preservation; instead, different designs exhibit complementary strengths that remain stable across datasets and neighborhood sizes. Our study provides a controlled and reproducible understanding of semantic ID design and offers practical insights for future generative recommender systems.\footnote{Our data and code are available at ~\url{https://github.com/layingfish/SID-Repro}.}

\end{abstract}


\received{20 February 2007}
\received[revised]{12 March 2009}
\received[accepted]{5 June 2009}

\maketitle
\section{Introduction}

Generative recommendation has become one of the most rapidly growing directions in recommender systems~\cite{geng2022recommendation,wang2023generative,li2024large,hou2025survey}. 
Unlike traditional retrieve-then-rank pipelines, generative recommendation formulates recommendation as a sequence generation problem, where the target item is generated autoregressively as a sequence of tokens. 
This paradigm naturally leverages the strong generative capabilities of modern transformer models and provides an end-to-end solution for recommendation. 
A key technique that enables this paradigm is the use of \emph{semantic IDs} (SIDs)~\cite{rajput2023recommender}, which replace arbitrary item identifiers with discrete codes derived from item semantics. Compared with randomly assigned IDs, SIDs carry richer semantic information and preserve meaningful relationships among items, making them more suitable for sequence generation. Moreover, by representing items with semantically grounded tokens, SIDs can better leverage the representational capacity and pre-trained knowledge of language models.
As a consequence, SIDs have become a mainstream design choice in recent generative recommendation methods.

However, despite the popularity of semantic IDs, their design is still far from unified.
Existing work has explored a broad range of semantic ID construction strategies, including various clustering-based, quantization-based, hierarchical, and learned discrete coding methods~\cite{wang2024eager,si2024generative,rajput2023recommender,wang2024learnable}.
While such diversity reflects the flexibility of semantic ID design, it also leads to a fundamental limitation in the current literature: there is still no systematic study that examines how different semantic ID designs affect generative recommendation under a unified experimental framework.
Since prior studies are often conducted with different datasets, backbones, training recipes, and implementation details, the community still lacks a clear understanding of which design choices are truly effective and which observed improvements are confounded by inconsistent settings. 
To bridge this gap, we conduct a large-scale reproducibility study on semantic ID design for generative recommendation. 
We ask a key research question: \emph{What makes a good semantic ID for generative
recommendation?}

While the primary objective of semantic ID design is to improve recommendation performance, its effectiveness can be evaluated from multiple complementary aspects, such as codebook utilization, scaling behavior, and item semantic understanding. To comprehensively answer this core question, we further decompose it into the following four research questions:

\noindent \textbf{RQ1. Which representative SID-based generative recommendation system performs best under a unified data split and evaluation protocol?}
This question focuses on a method-level benchmark where all methods share the same preprocessing, data split, and evaluation scorer, while retaining their original training and inference designs.

\noindent \textbf{RQ2. Which SID design has the best codebook utilization and does that lead to better recommendation performance?}  
Codebook utilization is commonly viewed as an indicator of semantic ID quality, but its connection to downstream recommendation performance remains unclear. 
We therefore examine whether higher utilization consistently translates into better recommendation results.

\noindent \textbf{RQ3. What are the scaling behaviors of generative recommendation based on different SIDs?}  
We investigate the scaling properties of SID-based recommendation from two key perspectives: the SID length and the size of the generative recommender backbone.
Specifically, we study how increasing SID length and enlarging the T5 model affect the representational capacity, generation difficulty, and overall recommendation performance.

\noindent \textbf{RQ4. How do different semantic ID designs affect local item semantic preservation?}
Since semantic IDs are intended to encode meaningful item structure, it is important to assess whether different designs preserve local semantic neighborhoods in the induced SID space.

By answering these research questions, we aim to move the study of semantic IDs from fragmented empirical practice toward a more principled and reproducible understanding. To this end, we build a unified experimental framework and evaluate representative SID-based generative recommendation methods on multiple benchmark datasets.
Our framework contains two levels of comparison: an overall method-level benchmark under a unified evaluation setup, and controlled SID-only analyses where the generative backbone and training protocol are fixed. 
Our results reveal that the effects of SID design are consistently non-monotonic. 
First, there is no universally best SID design: many methods are dataset-dependent, and even widely used RQ-VAE- and OPQ-based designs are not stable across all datasets. 
Second, better first-level codebook utilization does not consistently translate into better recommendation performance. 
Third, scaling is also non-monotonic: increasing either the generative backbone size or the SID length does not always improve results. 
Finally, across four datasets, OPQ and RQ-Kmeans exhibit complementary strengths in preserving local semantic structure: OPQ better recovers semantic neighborhoods, whereas RQ-Kmeans better preserves the ordering of the closest semantic neighbors.

Our contributions are summarized as follows:

\begin{itemize}
    \item We present a large-scale reproducibility study on semantic ID design for generative recommendation, combining an overall method benchmark with controlled SID-only analyses under a unified preprocessing and evaluation protocol.
    
    \item We complement ranking metrics with cross-dataset diagnostics of codebook balance and local semantic preservation, covering multiple data scales and neighborhood sizes.
    
    \item We provide practical insights into how to design effective semantic IDs for generative recommendation, showing which observations are stable across datasets and which remain dataset- or protocol-dependent.
\end{itemize}

\section{Related Work}

\noindent \textbf{Generative recommendation.}
Generative recommendation formulates recommendation as a conditional generation problem, where the target item is generated from user behavior sequences rather than selected only through a conventional retrieve-and-rank pipeline~\cite{geng2022recommendation,ji2024genrec,rajput2023recommender,deldjoo2024recommendation}.
Early text-to-text methods such as P5 unify different recommendation tasks with personalized prompts, while recent LLM-based recommenders further explore item generation, instruction tuning, and language-model adaptation for recommendation~\cite{geng2022recommendation,ji2024genrec,bao2023tallrec}.
Although these methods demonstrate the flexibility of generation-based recommendation, they also reveal a fundamental target-representation challenge.
Generating item titles or textual descriptions can introduce ambiguity, synonymy, and unnecessary language-generation difficulty, whereas generating atomic item IDs avoids textual ambiguity but provides little semantic sharing across related items.
This motivates the use of structured item identifiers as generation targets.
A representative method is TIGER, which introduces semantic IDs by mapping each item into a sequence of discrete codes and generating the target item identifier autoregressively~\cite{rajput2023recommender}.
Since then, semantic IDs have become a central interface between item representation and generative recommendation models.

\noindent \textbf{Semantic ID design.}
Following TIGER, existing SID methods can be roughly organized by how they construct the discrete item code and how they incorporate collaborative information~\cite{penha2025semantic,fu2026differentiable,fu2025forge,fang2025hid,zhang2026cold,chenstar,baikalov2026mitigating,valeau2026efficient,wang2026momorec,khrylchenko2026variable,liu2024mmgrec,hou2025towards}.
From the construction perspective, TIGER uses RQ-VAE codes, while LETTER and ETEGRec further refine RQ-VAE-style tokenization with regularization or end-to-end learning~\cite{rajput2023recommender,wang2024learnable,liu2025generative}.
Other methods explore different target spaces: RPG and DiffGRM adopt OPQ-based identifiers, OneRec uses RQ-Kmeans, HowToIndex relies on metadata/content-based indexing, EAGER builds hierarchical K-means behavior--semantic codes, SEATER constructs tree-structured SIDs, and SETRec uses an AE-based set tokenizer~\cite{hua2023index,hou2025generating,liu2026diffgrm,deng2025onerec,wang2024eager,si2024generative,lin2025order,liu2025onerec}.
This line shows a trend from simple autoregressive semantic codes toward more structured, parallel, hierarchical, and order-agnostic target spaces.
From the preference-signal perspective, early SIDs mainly rely on content semantics, while later methods increasingly inject CF information: LETTER uses CF as tokenization regularization, SEATER incorporates CF into tree construction, SETRec and EAGER combine semantic and CF representations, and ETEGRec jointly learns the tokenizer and recommender with recommendation alignment~\cite{wang2024learnable,si2024generative,lin2025order,wang2024eager,liu2025generative}.
Together, these studies show that SID design shapes not only the target space, but also semantic sharing, collaborative alignment, and decoding difficulty in generative recommendation.

The importance of SID design is further reflected in recent industrial generative recommendation systems.
Large-scale systems such as OneRec  show that generative recommendation is moving toward production deployment, while GLIDE and MBGR explicitly adopt semantic-ID-based or business-aware ID designs for large-scale retrieval and recommendation~\cite{deng2025onerec,li2026mbgr,d2026deploying,ju2026semantic}.
However, existing SID methods are usually evaluated under different datasets, preprocessing pipelines, model backbones, and decoding implementations, making it difficult to isolate the effect of the identifier itself.
Motivated by the growing importance of semantic IDs and the lack of systematic reproducible comparison, we treat SID design as the main object of study and evaluate representative SID designs under a unified framework.

\section{Preliminaries}
\noindent \textbf{Task formulation.}
Let $\mathcal{U}$ and $\mathcal{I}$ denote the sets of users and items, respectively. 
For a user $u\in\mathcal{U}$, let 
$S_u^t=(i_1^u,\ldots,i_t^u)$ denote the historical interaction sequence up to time $t$. 
During training, each sequence prefix defines a next-interaction prediction instance with target $i_{t+1}^u\in\mathcal{I}$. During evaluation, each user is evaluated once at the global cutoff, with all subsequent warm interactions treated as relevant items.

In semantic-ID-based generative recommendation, items are represented by sequences of discrete tokens instead of atomic item identifiers. 
A semantic-ID tokenizer 
$q:\mathcal{I}\rightarrow \mathcal{C}_1\times\cdots\times\mathcal{C}_L$ 
maps each item $i$ to an $L$-token code sequence,
\[
q(i)=(z_{i,1},\ldots,z_{i,L}),\quad z_{i,\ell}\in\mathcal{C}_{\ell}.
\]
Here $\mathcal{C}_{\ell}$ denotes the codebook at level $\ell$. 
The tokenizer $q$ can be constructed using clustering, vector quantization, product quantization, tree-structured assignment, or learnable tokenization.

Given a user history $S_u^t$, the recommender first converts each interacted item into its semantic ID, yielding
\[
Z_u^t=(q(i_1^u),\ldots,q(i_t^u)),
\]
where $Z_u^t$ denotes the semantic ID sequence of user $u$ up to time step $t$. The model then learns to generate the semantic ID of the target item. Assuming a one-to-one mapping between valid semantic IDs and items, the standard left-to-right autoregressive SID generation objective can be written as:
\[
p_{\theta}(i_{t+1}^u\mid S_u^t)
\equiv
p_{\theta}(q(i_{t+1}^u)\mid Z_u^t)
=
\prod_{\ell=1}^{L}
p_{\theta}
\left(
z_{i_{t+1}^u,\ell}
\mid
Z_u^t,
z_{i_{t+1}^u,<\ell}
\right),
\]
where $z_{i_{t+1}^u,<\ell}$ denotes the prefix tokens before level $\ell$.

At inference time, the model generates valid semantic-ID sequences and maps them back to items through a code-to-item dictionary. 
Existing methods mainly differ in how the tokenizer $q$ is designed and how the semantic ID is decoded. 
For example, some methods generate tokens from left to right, whereas others adopt constrained decoding, parallel prediction, or order-agnostic generation. 
The training objective is typically the negative log-likelihood of the target semantic ID, optionally combined with reconstruction, contrastive, diversity, or collaborative regularization losses.

Built upon this semantic-ID-based generative recommendation paradigm, this paper focuses on understanding how the design and properties of semantic IDs affect recommendation performance.

\section{Experiment Setup}
\begin{table}[t]
  \centering
  \caption{Statistics of the datasets used in our experiments.}
  \label{tab:dataset_statistics}
   \renewcommand{\arraystretch}{0.9} 
  \begin{tabular}{lrrr}
  \toprule
  Dataset & \#Users & \#Items & \#Interactions \\
  \midrule
  Video Games & 74,333 & 25,062 & 666,397 \\
  Microlens-50K & 42,666 & 14,079 & 310,531 \\
  Microlens-100K & 90,171 & 17,228 & 658,056 \\
  Yelp & 221,039 & 109,326 & 3,531,723 \\
  \bottomrule
  \end{tabular}
  \vspace{-0.2in}
  \end{table}
\subsection{Dataset}
We conduct experiments on multiple benchmark datasets, Amazon \textit{Video Games}\footnote{\url{https://amazon-reviews-2023.github.io}.}, \textit{Microlens}\footnote{\url{https://github.com/westlake-repl/MicroLens}.}, and \textit{Yelp}\footnote{\url{https://business.yelp.com/data/resources/open-dataset/}}, covering three different recommendation scenarios and data characteristics. The detailed dataset statistics are presented in Table~\ref{tab:dataset_statistics}. The Microlens dataset comprises two variants: we adopt Microlens-50K as the default benchmark, while Microlens-100K is used in the additional utilization and scaling analyses. RQ2 covers Video Games and both Microlens scales, and RQ4 covers all four datasets, including Yelp.
Following prior generative recommendation studies~\cite{rajput2023recommender,lin2025order,sun2020we}, we convert user--item interactions into chronological behavior sequences and adopt a unified global temporal split. Specifically, after applying the same filtering criteria to all methods, we merge all interactions into a global timestamp-ordered log and split it into training, validation, and test sets with a ratio of $70{:}13{:}17$. This ensures that validation and test interactions occur after the training period at the global level, reducing future information leakage. After splitting, each user's interactions are sorted chronologically to construct input sequences. Next interactions are used as training targets, while evaluation treats each user’s post-cutoff warm interactions as the ground-truth relevant set. The same preprocessing pipeline is used for all baselines so that performance differences can be attributed to model and semantic ID design rather than data construction.
\subsection{Evaluation}
For fair comparison, we use the same global temporal split for all datasets: after $5$-core filtering, interactions are sorted by timestamp and split by global time cutoffs, ensuring that test interactions occur after the training period~\cite{lin2025order}. At each evaluation cutoff, each user is evaluated once, with all subsequent warm items in the corresponding evaluation split treated as relevant. We therefore report multi-target Recall@5, Recall@10, NDCG@5, and NDCG@10, averaged uniformly across users. Our main protocol is warm-only: cold items are removed from the ground-truth set before evaluation, and users with no remaining ground-truth items are excluded. For Section~\ref{sec:rq1} (RQ1), each method is evaluated under its original inference setting. 
For the controlled analyses from Section~\ref{sec:utilization} (RQ2) to Section~\ref{sec:learning_dyanamics} (RQ4), we fix the TIGER framework and vary only the SID assignment. 
Specifically, we choose TIGER as the controlled backbone because it provides a simple and widely adopted autoregressive SID generation framework, allowing us to isolate the effect of the identifier assignment itself while keeping the training objective, decoding procedure, and evaluation pipeline unchanged.
Following commonly adopted generative recommendation approaches~\cite{rajput2023recommender,wang2024learnable}, we generate top-$K$ predictions using constrained trie-based prefix decoding in these controlled experiments. 
The resulting candidates are passed through a shared post-processing pipeline that removes items appearing in the user's training history and fills any remaining slots with training-time popular items, thereby guaranteeing $K$ valid recommendations. 
All metrics are computed with the same strict scorer, so RQ1 reflects complete-method performance, while RQ2--RQ4 mainly reflect the effect of SID design.

\subsection{Baselines}

To provide a comprehensive and fair comparison, we include baselines from both \textbf{conventional sequential recommendation} and \textbf{generative recommendation with semantic IDs}. These methods cover major SID construction strategies, including RQ-VAE, OPQ, RQ-Kmeans, metadata/content-based indexing, hierarchical clustering, tree-based identifiers, AE-based set tokenization, and end-to-end learnable tokenization.

\noindent \textbf{1. Conventional sequential recommendation.}
\begin{itemize}
    \item \textit{SASRec~\cite{kang2018self}} is a strong self-attention-based sequential recommendation baseline.
\end{itemize}

\noindent \textbf{2. Generative recommendation with semantic  IDs.}
\begin{itemize}
    \item \textit{TIGER~\cite{rajput2023recommender}} maps item semantics to RQ-VAE code sequences and generates the target SID autoregressively.

    \item \textit{HowToIndex~\cite{hua2023index}} represents metadata/content-based item indexing. It constructs item identifiers from explicit item-side information or predefined indexing rules, providing a comparison to learned SID methods.

    \item \textit{RPG~\cite{hou2025generating}} is an OPQ-based method for long semantic IDs. It constructs unordered long SIDs through optimized product quantization and predicts SID tokens in parallel.

    \item \textit{LETTER-TIGER~\cite{wang2024learnable, rajput2023recommender}} applies the LETTER tokenizer to the TIGER backbone. LETTER improves RQ-VAE tokenization with semantic regularization, CF-based contrastive regularization, and diversity regularization.

    \item \textit{LETTER-LC-Rec~\cite{wang2024learnable, zheng2024adapting}} applies the same LETTER tokenizer to the LC-Rec backbone, allowing us to examine the effect of CF-regularized SID construction across different generative backbones.

    \item \textit{SETRec~\cite{lin2025order}} adopts an order-agnostic set identifier paradigm. It represents each item with CF and semantic tokens, where the semantic component is produced by an AE-based semantic tokenizer.

    \item \textit{ETEGRec~\cite{liu2025generative}} is an end-to-end learnable item tokenization framework that jointly optimizes an RQ-VAE-based tokenizer and a T5-like generative recommender with recommendation alignment objectives.

    \item \textit{SEATER~\cite{si2024generative}} adopts a semantic tree-structured identifier design. It constructs balanced tree-based item identifiers with collaborative item embeddings and uses generation and contrastive learning objectives.

    \item \textit{EAGER~\cite{wang2024eager}} is a behavior--semantic two-stream generative recommender. It builds behavior and semantic codes through hierarchical K-means and generates them with two separate decoders.

    \item \textit{OneRec~\cite{deng2025onerec}} represents a RQ-Kmeans-based SID construction strategy. It uses multi-level balanced quantization to generate semantic ID tokens and unifies retrieval and ranking in a generative framework.

    \item \textit{DiffGRM~\cite{liu2026diffgrm}} is a diffusion-based generative recommendation model. It constructs SIDs with parallel semantic encoding and uses masked discrete diffusion with confidence-guided parallel denoising.
\end{itemize}

Overall, these baselines cover diverse SID construction strategies, decoding mechanisms, and CF incorporation designs, enabling a broad comparison of representative SID-based systems. 

\subsection{Implementation Details}
For all compared methods, we mainly follow the hyperparameter settings, training recipes, and official implementations reported in their original papers to ensure a reproducible comparison. For semantic-ID-based methods in Section~\ref{sec:rq1}, we keep the default tokenizer and codebook configurations from the corresponding original setups unless otherwise specified. From Section~\ref{sec:utilization} (RQ2) to Section~\ref{sec:learning_dyanamics} (RQ4), we focus on semantic ID design. To enable meaningful comparisons and obtain clearer insights into SID behavior, we fix the SID implementation\footnote{\url{https://github.com/EdoardoBotta/RQ-VAE-Recommender}} and the TIGER training protocol, varying only the SID design. Consequently, the performance of the same method may slightly differ between RQ1 and RQ2--RQ4.

For RQ2, we evaluate TIGER, RQ-Kmeans, LETTER-div, and LETTER-no-div on Video Games, Microlens-50K, and Microlens-100K. We compute used codes, entropy, normalized entropy, perplexity, effective usage, Gini coefficient, prefix utilization, and collision rate at all three levels; the main paper focuses on first-level normalized entropy because deeper levels are close to saturation. For RQ4, we evaluate six SID designs on Video Games, both Microlens scales, and Yelp using $K\in\{10,20,50\}$. In the code-length scaling experiments, we vary the number of semantic ID levels as $L \in \{2,3,4,6,8,12,16\}$ for RQ-VAE-, RQ-Kmeans-, and OPQ-based identifiers. For backbone scaling, we instantiate the generative recommender using pre-trained T5-small, T5-base, and T5-large checkpoints~\cite{raffel2020exploring}. All models are trained with early stopping based on validation NDCG@10, and the final results are reported on the test set using the same evaluation pipeline. Experiments are conducted on eight RTX 4090 GPUs.

\noindent \textbf{Evidence scope.} RQ1 prioritizes faithful reproduction over exhaustive dataset-specific tuning, so small method gaps should be interpreted cautiously. Our strongest claims rely on repeated patterns or controlled contrasts: RQ2 is supported by three datasets and the LETTER diversity ablation, RQ3 reports both data- and identifier-scale changes, and RQ4 covers four datasets and three neighborhood sizes. We therefore distinguish stable trends from rankings that may remain sensitive to tuning or backbone choice.

\begin{table*}[t]
\centering
\caption{Overall performance comparison under a unified data split and scorer, while preserving method-specific inference settings. The best-performing methods are shown in bold, and the second-best are underlined.}
 \vspace{-0.15in}
 \renewcommand{\arraystretch}{0.9} 
\label{tab:main_results}
\resizebox{\textwidth}{!}{
\begin{tabular}{lcccccccccccc}
\toprule
\multirow{2}{*}{Model} 
& \multicolumn{4}{c}{Video Games} 
& \multicolumn{4}{c}{Microlens} 
& \multicolumn{4}{c}{Yelp} \\
\cmidrule(lr){2-5} \cmidrule(lr){6-9} \cmidrule(lr){10-13}
& R@5 & R@10 & N@5 & N@10 
& R@5 & R@10 & N@5 & N@10 
& R@5 & R@10 & N@5 & N@10 \\
\midrule
SASRec        & 0.0171 & 0.0300 & 0.0108 & 0.0152 & 0.0179 & 0.0318 & 0.0109 & 0.0156 & 0.0136 & 0.0257 & 0.0136 & 0.0177 \\
TIGER         & \underline{0.0287} & \underline{0.0468} & \underline{0.0209} & \underline{0.0272} & 0.0099 & 0.0157 & 0.0069 & 0.0089 & 0.0138 & 0.0230 & 0.0144 & 0.0172 \\
HowToIndex     & 0.0244 & 0.0369 & 0.0190 & 0.0232 & 0.0078 & 0.0129 & 0.0053 & 0.0070 & 0.0066 & 0.0110 & 0.0074 & 0.0086 \\
RPG           & \textbf{0.0337} & \textbf{0.0477} & \textbf{0.0254} & \textbf{0.0301} & \underline{0.0193} & \underline{0.0333} & \underline{0.0126} & \underline{0.0173} & \underline{0.0169} & \underline{0.0287} & \underline{0.0169} & \underline{0.0207} \\
LETTER-TIGER  & 0.0277 & 0.0437 & 0.0203 & 0.0257 & 0.0131 & 0.0231 & 0.0088 & 0.0121 & \textbf{0.0186} & \textbf{0.0322} & \textbf{0.0187} & \textbf{0.0232} \\
LETTER-LC-Rec & 0.0103 & 0.0187 & 0.0077 & 0.0105 & 0.0108 & 0.0176 & 0.0068 & 0.0091 & 0.0142 & 0.0232 & 0.0135 & 0.0167 \\
SETRec        & 0.0167 & 0.0256 & 0.0128 & 0.0159 & 0.0127 & 0.0217 & 0.0087 & 0.0117 & 0.0157 & 0.0279 & 0.0162 & 0.0201 \\
ETEGRec       & 0.0191 & 0.0320 & 0.0145 & 0.0189 & 0.0187 & \underline{0.0333} & 0.0124 & \underline{0.0173} & 0.0157 & 0.0275 & 0.0159 & 0.0197 \\
SEATER        & 0.0206 & 0.0330 & 0.0142 & 0.0184 & \textbf{0.0221} & \textbf{0.0356} & \textbf{0.0143} & \textbf{0.0189} & 0.0154 & 0.0277 & 0.0162 & 0.0201 \\
EAGER         & 0.0205 & 0.0311 & 0.0147 & 0.0183 & 0.0142 & 0.0229 & 0.0097 & 0.0126 & 0.0142 & 0.0252 & 0.0147 & 0.0183 \\
DiffGRM       & 0.0149 & 0.0228 & 0.0101 & 0.0127 & 0.0117 & 0.0192 & 0.0076 & 0.0101 & 0.0134 & 0.0233 & 0.0142 & 0.0173 \\
OneRec        & 0.0219 & 0.0322 & 0.0156 & 0.0191 & 0.0172 & 0.0265 & 0.0114 & 0.0147 & 0.0118 & 0.0194 & 0.0131 & 0.0152 \\
\bottomrule
\end{tabular}
}
\end{table*}
\subsection{Overall Benchmark (RQ1)}
\label{sec:rq1}

To answer RQ1 and investigate the complete-system performance of representative SID-based methods, we conduct a large-scale benchmark on public datasets: Video Games, Microlens, and Yelp. We first analyze the main results in Table~\ref{tab:main_results}, and then further discuss two aspects: SID construction strategies and CF signal incorporation.

The main observation from Table~\ref{tab:main_results} is that no method dominates across all datasets. 
RPG is the strongest overall method, ranking first on Video Games and second on both Microlens and Yelp, but the best-performing method still varies by dataset: SEATER performs best on Microlens, while LETTER-TIGER performs best on Yelp. 
This dataset dependence is also visible for other methods, whose relative rankings change substantially across benchmarks. 
Meanwhile, SASRec remains a competitive lightweight baseline and outperforms several generative methods on Microlens and Yelp. 
This suggests that generative recommendation does not automatically outperform conventional sequential recommendation. The benefit depends critically on whether the SID design and the generative objective are well aligned with the data.

\begin{figure}[t]
    \centering
    \includegraphics[width=\linewidth]{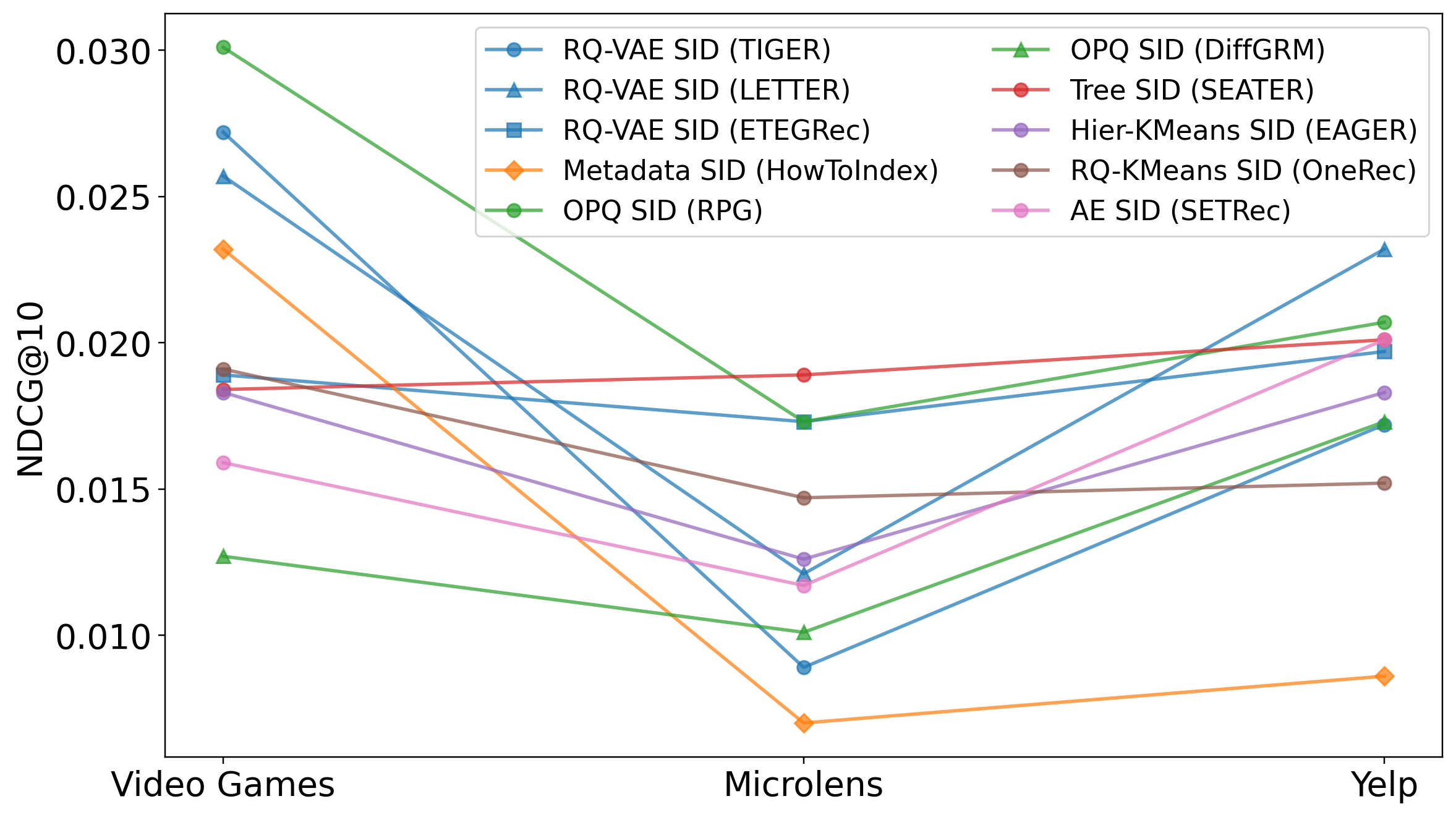}
    \caption{Dataset-wise NDCG@10 comparison of different SID-based methods on Video Games, Microlens, and Yelp. Each point denotes the result on one dataset, and lines connect the same SID design across datasets.}
    \label{fig:rq1_ndcg}
\end{figure}

From the perspective of SID construction, Figure~\ref{fig:rq1_ndcg} shows that the construction family alone is insufficient to explain recommendation quality. 
Methods within the same family can behave quite differently. 
For example, OPQ-based methods show a large gap between RPG and DiffGRM, and RQ-VAE-based methods also exhibit clear dataset-dependent behavior. 
This indicates that SID effectiveness is not determined only by the high-level tokenizer type, such as RQ-VAE, OPQ, tree-based indexing, or metadata-based indexing. 
Instead, the full system design, including the tokenizer objective, code organization, decoding strategy, and its compatibility with the recommendation data, jointly determines performance.

\begin{figure}[t]
    \centering
    \includegraphics[width=\linewidth]{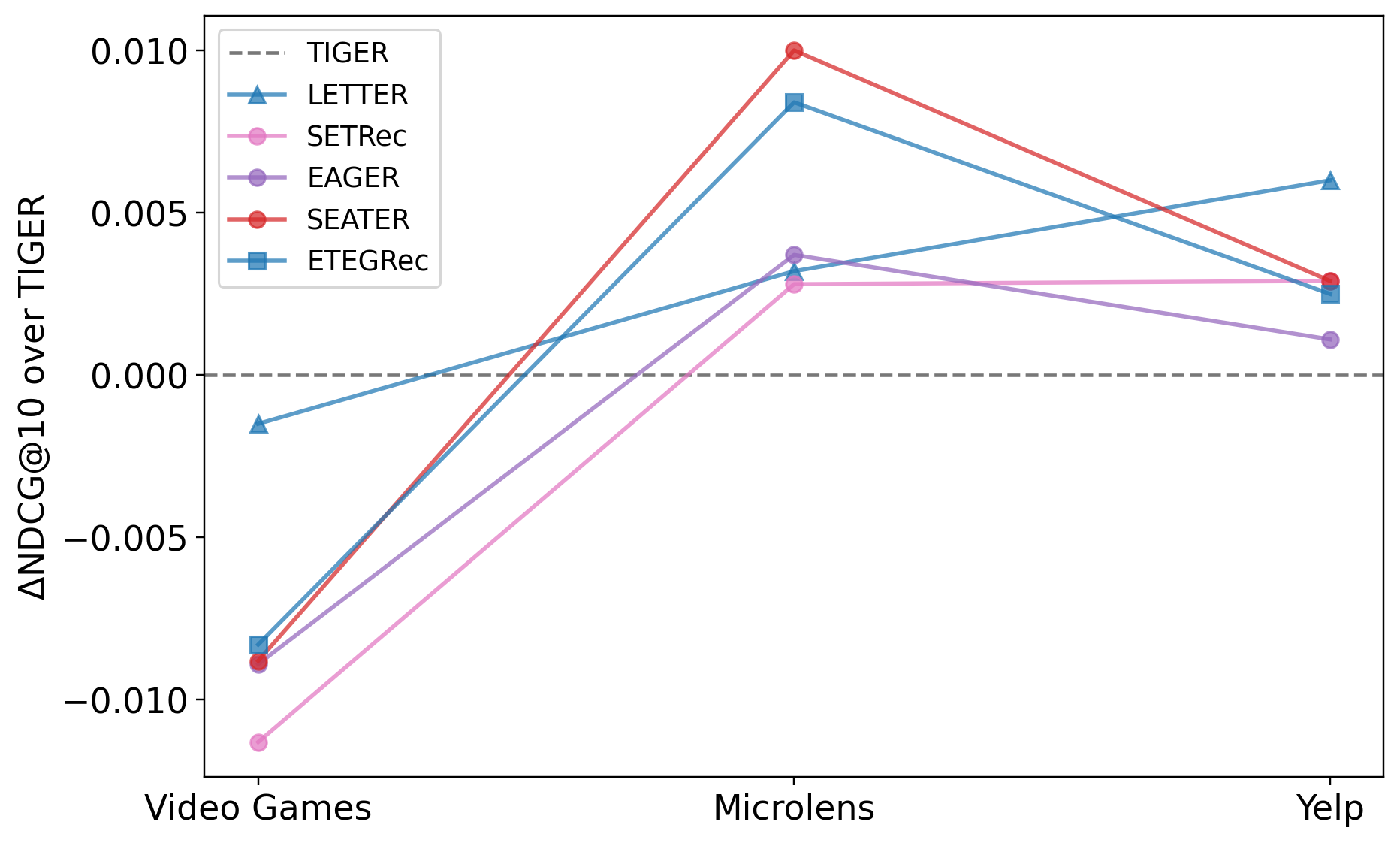}
    \caption{Dataset-wise $\Delta$NDCG@10 over TIGER for methods that incorporate collaborative filtering signals. Positive values indicate improvement over the baseline.}
     \vspace{-0.1in}
    \label{fig:rq1_cf_ndcg}
\end{figure}

From the CF signal exploitation perspective, Figure~\ref{fig:rq1_cf_ndcg} shows that incorporating collaborative information is not uniformly beneficial. 
On Video Games, most CF-enhanced methods underperform the TIGER baseline, whereas on Microlens and Yelp several CF-enhanced methods bring clear gains. 
This suggests that CF signals can improve SID quality when collaborative patterns are reliable and aligned with item semantics, but may introduce noise or instability when such alignment is weak. 
Among the CF-incorporating methods, LETTER-TIGER shows relatively stable behavior, suggesting that using CF as a regularization signal during SID construction may be a more robust way to inject collaborative information than directly relying on CF representations throughout the full generative pipeline.

\noindent \textbf{Answer to RQ1:} \textbf{The key conclusion is that SID design substantially affects generative recommendation performance, but the best design varies across datasets.} This variation suggests that SID design should be adapted to different recommendation scenarios rather than treated as a universal choice. Different item semantics and user behavior patterns may require different identifier structures. First, RPG achieves the best overall benchmark performance, ranking first on Video Games and second on both Microlens and Yelp. Second, SASRec remains a strong baseline against generative methods on some datasets. Third, there is no universally best SID construction strategy; even widely used OPQ and RQ-VAE designs show large performance variation. Fourth, CF signal incorporation is also dataset-dependent. Among CF-incorporating methods, LETTER-TIGER shows the most stable behavior, suggesting that using CF information as a regularization signal during SID construction may be a relatively robust way to incorporate collaborative information.

\section{Codebook Utilization Investigation (RQ2)}
\label{sec:utilization}

\begin{figure*}[t]
    \centering
    \includegraphics[width=0.8\textwidth]{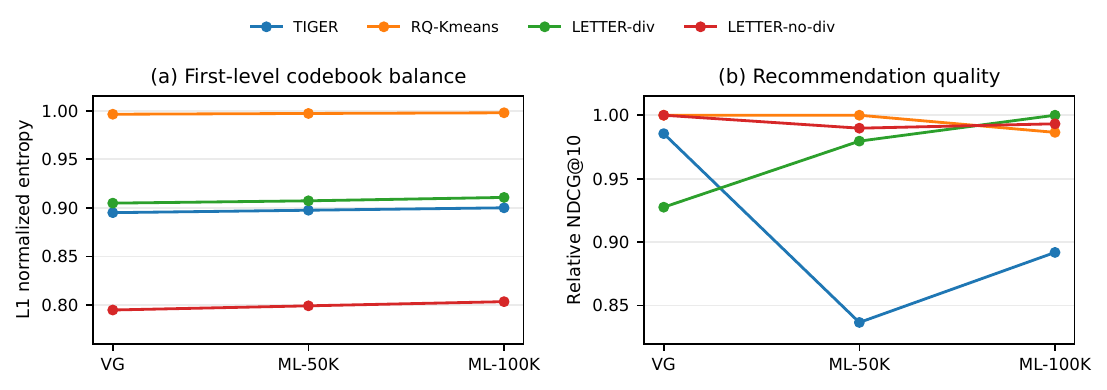}
    \caption{Cross-dataset comparison of codebook balance and recommendation quality under the controlled TIGER protocol. Left: first-level normalized entropy, where larger values indicate more balanced code use. Right: NDCG@10 divided by the best value within each dataset, so $1$ denotes the dataset-wise best result. RQ-Kmeans is consistently the most balanced tokenizer, but the recommendation ranking does not follow the same ordering.}
    \Description{Two line charts compare four semantic-ID designs on Video Games, Microlens-50K, and Microlens-100K. RQ-Kmeans has the highest normalized entropy on every dataset, while methods with substantially lower entropy can match or exceed its relative NDCG at some datasets.}
    \label{fig:rq2_utilization_cross_dataset}
\end{figure*}

\begin{table*}[t]
\centering
\caption{Complete codebook diagnostics for RQ2 under the controlled TIGER protocol. Each slash-separated entry reports the values at L1/L2/L3. $H$ is entropy, $H^{\mathrm{norm}}$ is normalized entropy, PPL is perplexity, Eff. is effective usage, Prefix is prefix utilization, and Coll. is the item-level collision rate. Values are rounded to two decimals.}
\label{tab:rq2_full_diagnostics}
\setlength{\tabcolsep}{2.2pt}
\renewcommand{\arraystretch}{0.82}
\begin{tabular}{llcccccccc}
\toprule
Dataset & SID & Used & $H$ & $H^{\mathrm{norm}}$ & PPL & Eff. & Gini & Prefix & Coll. \\
\midrule
\multirow{4}{*}{VG}
& TIGER         & 161/252/256 & 4.96/5.47/5.52 & 0.90/0.99/1.00 & 143.00/237.00/248.50 & 0.56/0.93/0.97 & 0.43/0.10/0.05 & 0.63/0.78/0.98 & 0.02 \\
& RQ-Kmeans     & 256/256/256 & 5.53/5.52/5.52 & 1.00/1.00/1.00 & 251.00/250.00/249.00 & 0.98/0.98/0.97 & 0.03/0.04/0.04 & 1.00/0.75/0.99 & 0.02 \\
& LETTER-div    & 170/254/256 & 5.02/5.48/5.52 & 0.91/0.99/1.00 & 151.00/240.00/249.50 & 0.59/0.94/0.98 & 0.39/0.09/0.05 & 0.66/0.72/0.99 & 0.01 \\
& LETTER-no-div & 98/248/256  & 4.41/5.44/5.51 & 0.80/0.98/0.99 & 82.00/231.00/247.00  & 0.32/0.90/0.97 & 0.65/0.13/0.06 & 0.38/0.55/0.98 & 0.03 \\
\midrule
\multirow{4}{*}{M50}
& TIGER         & 164/253/256 & 4.98/5.47/5.52 & 0.90/0.99/1.00 & 145.00/238.00/249.00 & 0.57/0.93/0.97 & 0.42/0.10/0.05 & 0.64/0.79/0.99 & 0.02 \\
& RQ-Kmeans     & 256/256/256 & 5.53/5.53/5.52 & 1.00/1.00/1.00 & 252.00/251.00/250.00 & 0.98/0.98/0.98 & 0.03/0.04/0.04 & 1.00/0.75/0.99 & 0.01 \\
& LETTER-div    & 172/254/256 & 5.03/5.49/5.52 & 0.91/0.99/1.00 & 153.00/241.00/250.50 & 0.60/0.94/0.98 & 0.38/0.09/0.04 & 0.67/0.73/0.99 & 0.01 \\
& LETTER-no-div & 101/249/256 & 4.43/5.45/5.51 & 0.80/0.98/0.99 & 84.00/232.00/248.00  & 0.33/0.91/0.97 & 0.64/0.12/0.05 & 0.40/0.56/0.98 & 0.02 \\
\midrule
\multirow{4}{*}{M100}
& TIGER         & 166/254/256 & 4.99/5.48/5.52 & 0.90/0.99/1.00 & 147.00/240.00/250.00 & 0.57/0.94/0.98 & 0.41/0.09/0.05 & 0.65/0.79/0.98 & 0.02 \\
& RQ-Kmeans     & 256/256/256 & 5.53/5.53/5.53 & 1.00/1.00/1.00 & 253.00/252.00/251.00 & 0.99/0.98/0.98 & 0.03/0.03/0.04 & 1.00/0.76/0.99 & 0.02 \\
& LETTER-div    & 174/255/256 & 5.05/5.49/5.53 & 0.91/0.99/1.00 & 156.00/243.00/251.50 & 0.61/0.95/0.98 & 0.37/0.08/0.04 & 0.68/0.74/0.99 & 0.01 \\
& LETTER-no-div & 103/250/256 & 4.45/5.46/5.52 & 0.80/0.98/1.00 & 86.00/234.00/249.00  & 0.34/0.91/0.97 & 0.63/0.11/0.05 & 0.40/0.57/0.98 & 0.02 \\
\bottomrule
\end{tabular}%
\end{table*}

\begin{figure*}[t]
    \centering
    \includegraphics[width=\linewidth]{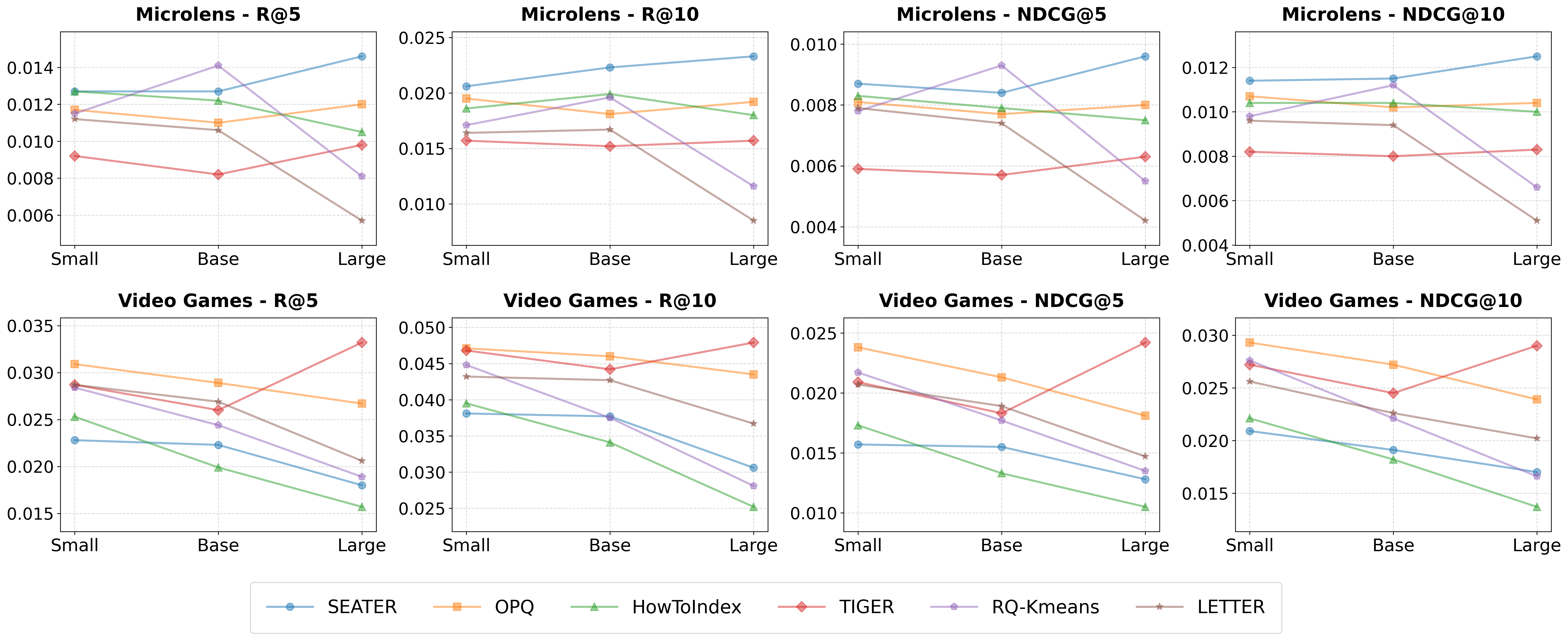}
    \caption{
    Scaling trends of different methods under small, base, and large T5 backbone on Video Games and Microlens. 
    Each curve corresponds to one method, showing how its ranking performance changes as the model scale increases.
    }
    \label{fig:ndcg_scaling}
\end{figure*}
\begin{table*}[t]
\centering
\caption{Performance comparison under small, base, and large backbones of pre-trained T5 on Video Games and Microlens. The best and second-best results for each dataset, metric, and backbone are marked in bold and underlined, respectively.}
\label{tab:ranking_results_combined}
\renewcommand{\arraystretch}{0.8} 
\resizebox{\textwidth}{!}{
\begin{tabular}{llcccccccccccc}
\toprule
\multirow{2}{*}{Dataset} & \multirow{2}{*}{SID Design} 
& \multicolumn{3}{c}{$R@5$}
& \multicolumn{3}{c}{$R@10$}
& \multicolumn{3}{c}{$N@5$}
& \multicolumn{3}{c}{$N@10$} \\
\cmidrule(lr){3-5} \cmidrule(lr){6-8} \cmidrule(lr){9-11} \cmidrule(lr){12-14}
& 
& small & base & large
& small & base & large
& small & base & large
& small & base & large \\
\midrule
\multirow{6}{*}{Microlens}
& TIGER      & 0.0092 & 0.0082 & 0.0098 & 0.0157 & 0.0152 & 0.0157 & 0.0059 & 0.0057 & 0.0063 & 0.0082 & 0.0080 & 0.0083 \\
& SEATER     & \textbf{0.0127} & \underline{0.0127} & \textbf{0.0146} & \textbf{0.0206} & \textbf{0.0223} & \textbf{0.0233} & \textbf{0.0087} & \underline{0.0084} & \textbf{0.0096} & \textbf{0.0114} & \textbf{0.0115} & \textbf{0.0125} \\
& OPQ    & \underline{0.0117} & 0.0110 & \underline{0.0120} & \underline{0.0195} & 0.0181 & \underline{0.0192} & 0.0081 & 0.0077 & \underline{0.0080} & \underline{0.0107} & 0.0102 & \underline{0.0104} \\
& HowToIndex & \textbf{0.0127} & 0.0122 & 0.0105 & 0.0186 & \underline{0.0199} & 0.0180 & \underline{0.0083} & 0.0079 & 0.0075 & 0.0104 & 0.0104 & 0.0100 \\
& LETTER     & 0.0112 & 0.0106 & 0.0057 & 0.0164 & 0.0167 & 0.0085 & 0.0079 & 0.0074 & 0.0042 & 0.0096 & 0.0094 & 0.0051 \\
& RQ-Kmeans     & 0.0115 & \textbf{0.0141} & 0.0081 & 0.0171 & 0.0196 & 0.0116 & 0.0078 & \textbf{0.0093} & 0.0055 & 0.0098 & \underline{0.0112} & 0.0066 \\
\midrule
\multirow{6}{*}{Video Games}
& TIGER      & \underline{0.0287} & 0.0260 & \textbf{0.0332} & \underline{0.0468} & \underline{0.0442} & \textbf{0.0479} & 0.0209 & 0.0183 & \textbf{0.0242} & 0.0272 & \underline{0.0245} & \textbf{0.0290} \\
& SEATER     & 0.0228 & 0.0223 & 0.0180 & 0.0381 & 0.0377 & 0.0306 & 0.0157 & 0.0155 & 0.0128 & 0.0209 & 0.0191 & 0.0170 \\
& OPQ    & \textbf{0.0309} & \textbf{0.0289} & \underline{0.0267} & \textbf{0.0471} & \textbf{0.0460} & \underline{0.0435} & \textbf{0.0238} & \textbf{0.0213} & \underline{0.0181} & \textbf{0.0293} & \textbf{0.0272} & \underline{0.0239} \\
& HowToIndex & 0.0253 & 0.0199 & 0.0157 & 0.0395 & 0.0341 & 0.0252 & 0.0173 & 0.0133 & 0.0105 & 0.0221 & 0.0182 & 0.0137 \\
& LETTER     & \underline{0.0287} & \underline{0.0269} & 0.0206 & 0.0432 & 0.0427 & 0.0367 & 0.0207 & \underline{0.0189} & 0.0147 & 0.0256 & 0.0226 & 0.0202 \\
& RQ-Kmeans     & 0.0284 & 0.0244 & 0.0189 & 0.0448 & 0.0375 & 0.0281 & \underline{0.0217} & 0.0177 & 0.0135 & \underline{0.0276} & 0.0221 & 0.0166 \\
\bottomrule
\end{tabular}}
\end{table*}

To answer RQ2, we conduct the utilization analysis on Video Games, Microlens-50K and Microlens-100K. We compare four directly compatible three-level variants: TIGER (RQ-VAE), RQ-Kmeans, LETTER with diversity regularization (LETTER-div), and LETTER without diversity regularization (LETTER-no-div). Each level contains 256 possible codes. Table~\ref{tab:rq2_full_diagnostics} reports the complete diagnostics, including used codes, entropy, normalized entropy, perplexity, effective usage, Gini coefficient, prefix utilization, and collision rate. For a concise cross-dataset comparison, Figure~\ref{fig:rq2_utilization_cross_dataset} uses first-level normalized entropy,
$H^{\mathrm{norm}}_1=-\sum_c p_{1,c}\log p_{1,c}/\log|\mathcal{C}_1|$,
where a value of $1$ indicates perfectly balanced code use.

Table~\ref{tab:rq2_full_diagnostics} shows a stable cross-dataset pattern. RQ-Kmeans achieves the most balanced L1 usage, with normalized entropy near $1.0$, high effective usage, and low Gini, whereas LETTER-no-div shows the strongest L1 imbalance. Differences largely vanish at L2/L3, where normalized entropy exceeds $0.98$ for all methods and collision rates remain low ($0.013$--$0.025$). However, utilization does not determine recommendation quality: RQ-Kmeans is not consistently best, and diversity regularization improves L1 normalized entropy by about $0.11$ while changing NDCG@10 by at most $0.002$. Across all settings, L1 normalized entropy has near-zero correlation with NDCG@10 (Pearson $r=-0.02$; Spearman $\rho=-0.08$).

The deeper levels provide little additional separation: L2 and L3 normalized entropy remains above $0.98$ for every setting, while collision rates stay low, between $0.013$ and $0.025$. Thus, the major structural difference lies in the coarse L1 partition, but even a substantially more balanced L1 partition does not guarantee better ranking quality. Utilization measures whether the tokenizer occupies its code space, not whether the resulting identifiers are semantically coherent and easy for the recommender to generate.

\noindent \textbf{Answer to RQ2:}
\textbf{Higher codebook utilization does not consistently lead to better recommendation performance, and this conclusion holds across Video Games and both Microlens scales}. RQ-Kmeans is consistently the most balanced tokenizer, yet lower-utilization variants can match or outperform it. Codebook utilization is therefore a useful diagnostic of tokenizer occupancy, but an insufficient standalone criterion for SID quality.

\begin{table}[t]
\centering
\caption{Performance comparison of different quantization methods under varying numbers of quantization levels $L$ on Video Games. For each method, the best result in each metric is highlighted in \textbf{bold}, and the second-best result is \underline{underlined}.}
 \vspace{-0.15in}
\label{tab:code_length}
\renewcommand{\arraystretch}{0.8} 
\begin{tabular}{lccccc}
\toprule
SID Design & $L$ & $R@5$ & $R@10$ & $N@5$ & $N@10$ \\
\midrule
RQ-VAE    & 2  & 0.0262 & 0.0448 & 0.0182 & 0.0255 \\
RQ-VAE    & 3  & \textbf{0.0287} & \underline{0.0468} & \textbf{0.0209} & \underline{0.0272} \\
RQ-VAE    & 4  & 0.0260 & 0.0409 & \underline{0.0203} & 0.0255 \\
RQ-VAE    & 6  & 0.0256 & 0.0440 & 0.0187 & 0.0248 \\
RQ-VAE    & 8  & \underline{0.0270} & 0.0458 & 0.0184 & 0.0257 \\
RQ-VAE    & 12 & 0.0266 & \textbf{0.0511} & 0.0194 & \textbf{0.0277} \\
RQ-VAE    & 16 & 0.0261 & 0.0424 & 0.0188 & 0.0244 \\
\midrule
RQ-Kmeans & 2  & 0.0270 & \underline{0.0443} & 0.0200 & 0.0252 \\
RQ-Kmeans & 3  & \textbf{0.0284} & \textbf{0.0448} & \textbf{0.0217} & \textbf{0.0276} \\
RQ-Kmeans & 4  & 0.0240 & 0.0369 & 0.0188 & 0.0233 \\
RQ-Kmeans & 6  & 0.0252 & 0.0364 & 0.0205 & 0.0241 \\
RQ-Kmeans & 8  & \underline{0.0279} & 0.0416 & \underline{0.0208} & \underline{0.0255} \\
RQ-Kmeans & 12 & 0.0214 & 0.0297 & 0.0179 & 0.0207 \\
RQ-Kmeans & 16 & 0.0267 & 0.0400 & 0.0199 & 0.0244 \\
\midrule
OPQ       & 2  & 0.0292 & 0.0469 & 0.0208 & 0.0267 \\
OPQ       & 3  & 0.0266 & 0.0413 & 0.0203 & 0.0253 \\
OPQ       & 4  & 0.0320 & \underline{0.0525} & 0.0230 & \underline{0.0300} \\
OPQ       & 6  & \textbf{0.0339} & \textbf{0.0556} & \textbf{0.0247} & \textbf{0.0321} \\
OPQ       & 8  & \underline{0.0336} & 0.0447 & \underline{0.0246} & 0.0284 \\
OPQ       & 12 & 0.0301 & 0.0433 & 0.0231 & 0.0275 \\
OPQ       & 16 & 0.0242 & 0.0354 & 0.0188 & 0.0226 \\
\bottomrule
\end{tabular}
\vspace{-0.1in}
\end{table}

\begin{figure}[t]
    \centering
    \includegraphics[width=\linewidth]{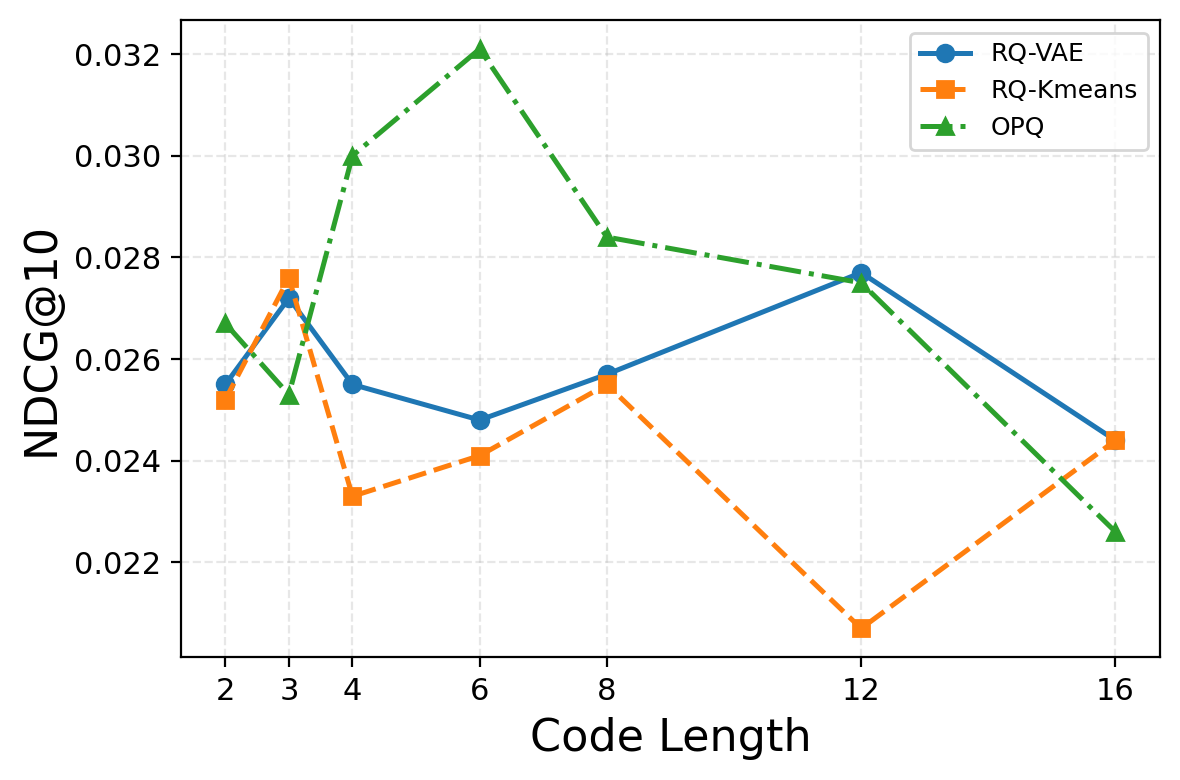}
    \caption{Scaling performance of different quantization methods with respect to code length on Video Games, evaluated by NDCG@10.}
    \label{fig:ndcg10_code_length}
\end{figure}

\section{Scaling Analysis (RQ3)}
\label{sec:scaling}
To answer RQ3, we focus on controlled SID design scaling analysis.
Among the compared SID designs, OPQ denotes the SID construction used by DiffGRM and RPG, while RQ-Kmeans denotes the SID construction used by OneRec.
Other SID constructions have been discussed in RQ1 as complete method-level designs and are therefore referred to by their method names, such as SEATER, HowToIndex, LETTER, and TIGER.
Under this controlled setting, we examine backbone scaling with T5-small/base/large and SID length scaling by varying the number of quantization levels $L$.
\begin{table}[t]
\centering
\caption{Additional backbone scaling results on larger Microlens-100K. The best result for each method and metric is highlighted in bold.}
\label{tab:microlens100k_backbone_scaling}
 \vspace{-0.15in}
\renewcommand{\arraystretch}{0.95}
\setlength{\tabcolsep}{3.5pt}
\begin{tabular}{lccccc}
\toprule
\textbf{SID Design} & \textbf{Backbone} & \textbf{R@5} & \textbf{N@5} & \textbf{R@10} & \textbf{N@10} \\
\midrule
\multirow{3}{*}{LETTER}
& T5-small & 0.0168 & 0.0113 & 0.0267 & 0.0148 \\
& T5-base  & \textbf{0.0219} & \textbf{0.0137} & \textbf{0.0381} & \textbf{0.0191} \\
& T5-large & 0.0206 & 0.0134 & 0.0357 & 0.0186 \\
\midrule
\multirow{3}{*}{RQ-Kmeans}
& T5-small & 0.0177 & 0.0117 & 0.0262 & 0.0146 \\
& T5-base  & \textbf{0.0215} & \textbf{0.0153} & \textbf{0.0352} & \textbf{0.0199} \\
& T5-large & 0.0208 & 0.0151 & 0.0330 & 0.0192 \\
\bottomrule
\end{tabular}
\vspace{-0.15in}
\end{table}

As shown in Figure~\ref{fig:ndcg_scaling} and Table~\ref{tab:ranking_results_combined}, backbone scaling exhibits non-monotonic behavior across datasets and SID designs. 
Moving from T5-small to T5-large does not consistently improve performance. 
Some methods remain relatively stable, but others degrade substantially under larger backbones, especially on Video Games. 
This suggests that larger generative capacity alone is insufficient for SID-based recommendation, and may even amplify optimization instability or overfitting when the identifier design is not well matched to the data. Although our study only scales up to T5-large due to computational constraints, the results still reveal an important trend: increasing backbone capacity alone does not necessarily guarantee better SID-based recommendation performance.

To further examine whether this instability is related to data scale, we conduct additional experiments on Microlens-100K, as shown in Table~\ref{tab:microlens100k_backbone_scaling}. 
For the two examined SID variants, the degradation from T5-base to T5-large becomes milder on the larger dataset, and T5-base provides clear gains over T5-small. 
This indicates that data sparsity may partly contribute to unstable backbone scaling, although the trend remains dependent on the specific SID design. 
Therefore, the observed degradation should be interpreted as practical scaling instability under a controlled training recipe, rather than as evidence that larger backbones are inherently worse.

Table~\ref{tab:code_length} and Figure~\ref{fig:ndcg10_code_length} further show that SID length scaling is also non-monotonic. 
Different quantization methods prefer different identifier lengths: RQ-based designs generally favor compact codes, while OPQ benefits from moderately longer codes but degrades when the identifier becomes too long. 
This pattern reflects a trade-off between identifier capacity and generation difficulty. 
Longer SIDs can reduce collisions and provide finer item separation, but they also lengthen the autoregressive prediction path and enlarge the sparse valid-code space, making exact identifier generation harder.

\noindent \textbf{Answer to RQ3:}
\textbf{The key conclusion is that SID scaling is non-monotonic in both backbone size and code length.}
First, larger backbones do not consistently help; degradation is more evident on Video Games for several methods, while Microlens shows mixed behavior. 
Second, the additional Microlens-100K results show that this degradation can be partially alleviated with more data, suggesting that overfitting under limited data is one possible cause.
Third, longer SIDs are not always better: RQ-Kmeans clearly favors compact codes around L=3, while RQ-VAE favors L=3 for top-5 metrics but reaches its best top-10 performance at L=12. OPQ performs best at a length L=6.
Overall, effective SID scaling requires balancing backbone capacity, identifier capacity, and data scale rather than simply increasing model size or code length.

\begin{figure*}[t]
    \centering
    \includegraphics[width=0.8\linewidth]{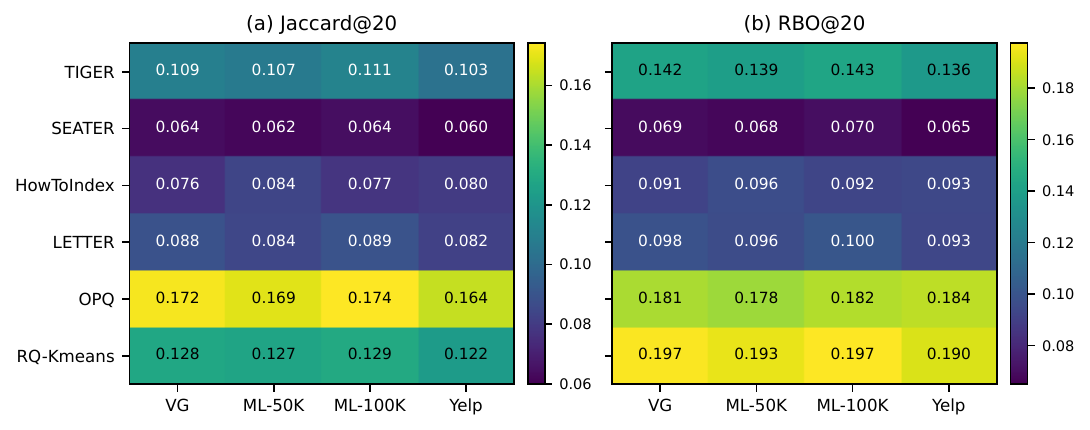}
    \caption{Local semantic preservation at $K=20$: OPQ leads Jaccard and RQ-Kmeans leads RBO on every dataset.}
    \Description{Two heat maps compare six semantic-ID designs on Video Games, Microlens-50K, Microlens-100K, and Yelp. OPQ is the strongest row in the Jaccard heat map, while RQ-Kmeans is the strongest row in the RBO heat map.}
    \label{fig:rq4_cross_dataset}
\end{figure*}
\begin{figure}[t]
    \centering
    \includegraphics[width=\linewidth]{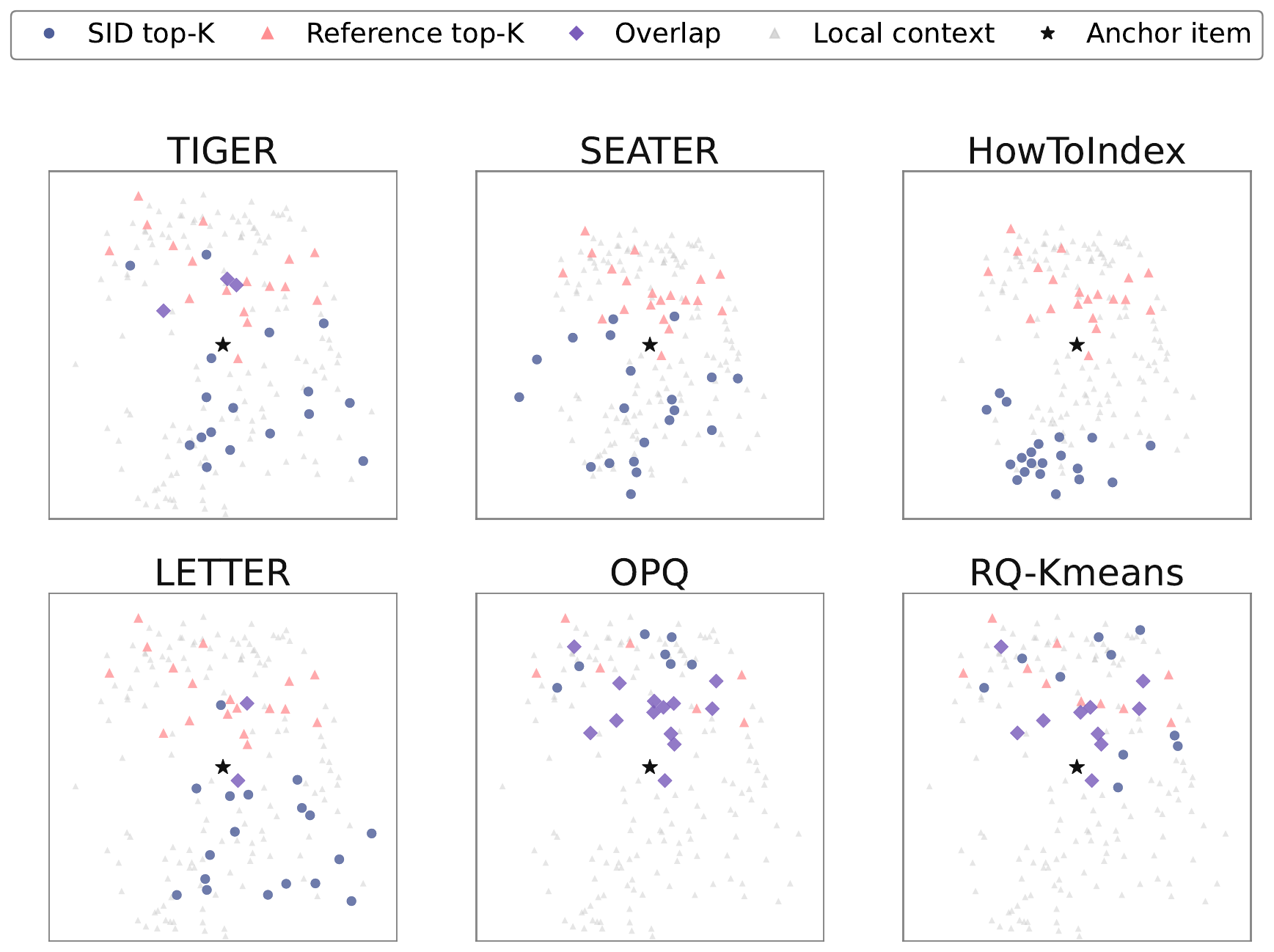}
    \caption{Anchor-level semantic neighborhoods on Video Games; purple markers denote reference--SID overlap.}
    \Description{Six two-dimensional neighborhood plots compare the reference and SID-induced nearest neighbors around the same anchor item. OPQ and RQ-Kmeans show visibly more overlap with the reference neighborhood than several other designs.}
    \vspace{-0.2in}
    \label{fig:rq4_semantic}
\end{figure}

\section{Local Semantic Preservation Analysis (RQ4)}
\label{sec:learning_dyanamics}

To answer RQ4, we conduct local semantic-preservation analysis on Video Games, Microlens-50K, Microlens-100K, and Yelp. For each item, we retrieve its top-$K$ nearest neighbors in the reference semantic space and under each SID design, and compare the two lists using Jaccard@$K$ and Rank-Biased Overlap (RBO)@$K$~\cite{webber2010similarity}. Jaccard measures strict neighbor-set recovery, whereas RBO gives larger weight to agreement near the top of the ranked lists. We evaluate $K\in\{10,20,50\}$ and report the main $K=20$ results in Figure~\ref{fig:rq4_cross_dataset}.

The cross-dataset pattern is stable. OPQ achieves the highest Jaccard@20 on all four datasets, with values from $0.1640$ to $0.1741$, indicating the strongest recovery of the reference neighbor set. RQ-Kmeans achieves the highest RBO@20 on all four datasets, with values from $0.1900$ to $0.1974$, indicating stronger preservation of the ordering among the closest neighbors. The gap also reflects two distinct geometric properties: OPQ exceeds RQ-Kmeans in Jaccard@20 by $0.0415$--$0.0448$, whereas RQ-Kmeans exceeds OPQ in RBO@20 by $0.0060$--$0.0161$.

The conclusion is also robust to neighborhood size. OPQ remains the best Jaccard method for every dataset at $K=10$, $20$, and $50$. RQ-Kmeans is the best RBO method on all datasets at $K=20$ and $K=50$, and on three of four datasets at $K=10$; the only reversal is Yelp at $K=10$, where OPQ exceeds RQ-Kmeans by just $0.0004$. Figure~\ref{fig:rq4_semantic} complements these aggregate results with an anchor-level example on Video Games. By contrast, SEATER consistently obtains the lowest values under both metrics, while TIGER, HowToIndex, and LETTER form a middle or lower tier. This shows that producing valid discrete identifiers does not necessarily preserve fine-grained semantic neighborhoods.

\noindent \textbf{Answer to RQ4:} \textbf{The cross-dataset results support a consistent conclusion while distinguishing two notions of semantic preservation.} OPQ consistently recovers the largest semantic-neighbor set, whereas RQ-Kmeans generally preserves the top-ranked neighbor order most strongly. This distinction holds across four datasets and is largely stable across $K\in\{10,20,50\}$.

\subsection{Practical Implications and Scope}
Three practices follow from the results. RPG is a strong complete-system starting point, but it should be checked against a competitive conventional baseline. Utilization statistics are useful for detecting collapse or imbalance, not as standalone optimization targets: LETTER's diversity regularization greatly improves balance without reliably improving ranking. Backbone size and SID length should be selected jointly on validation data, since the largest setting is often suboptimal; compact RQ codes and moderately long OPQ codes are safer defaults. A practical evaluation sequence is therefore to compare complete systems first, inspect utilization and collisions only as failure diagnostics, and then use both Jaccard and RBO when semantic transfer matters, because set recovery and rank preservation favor different tokenizers. These recommendations are conditioned on the study's scope: RQ2--RQ4 fix TIGER, RQ1 mainly follows official hyperparameters, and evaluation is warm-only, so exact orderings still require validation under other backbones, catalogs, and cold-start protocols.

\section{Conclusion and Future Work}
This paper studies what makes a good semantic ID through a unified benchmark and controlled SID-only analysis. Complete-system performance is dataset-dependent, although RPG is consistently strong. Under the fixed TIGER protocol, RQ-Kmeans has the most balanced first-level codebook but is not consistently the best recommender; larger backbones and longer SIDs are also not monotonically beneficial. Across four datasets, OPQ best recovers semantic-neighbor sets, while RQ-Kmeans best preserves their top-ranked order. Effective SID design should therefore be judged jointly by recommendation quality, utilization, generability, scaling behavior, and semantic geometry.

Future work should test whether these patterns survive under other generative backbones and decoding rules, systematic dataset-specific tuning, larger item catalogs, and cold-start evaluation. Error analysis should also connect prefix-level generation failures, candidate fallback, and optimization instability to the geometry of the learned identifier space.

\balance
\bibliographystyle{ACM-Reference-Format}
\bibliography{sample-base}


\end{document}